%% file: main.tex
\documentclass[nonacm]{acmart}

\usepackage{booktabs} 
\usepackage{algorithm}
\usepackage{algpseudocode}
\usepackage{times}
\usepackage{epsfig}
\usepackage{graphicx}
\usepackage{amsmath}
\usepackage{amssymb}

\begin{document}


\title{Deep Landscape Features for Improving Vector-borne Disease Prediction}


\author{Nabeel Abdur Rehman}
\affiliation{%
  \institution{New York University}
}
\email{nabeel@nyu.edu}

\author{Umar Saif}
\affiliation{%
  \institution{Information Technology University, Pakistan}
}
\email{umar@csail.mit.edu}

\author{Rumi Chunara}
\affiliation{%
  \institution{New York University}
}
\email{rumi.chunara@nyu.edu}


\begin{abstract}
The global population at risk of mosquito-borne diseases such as dengue, yellow fever, chikungunya and Zika is expanding. Infectious disease models commonly incorporate environmental measures like temperature and precipitation. Given increasing availability of high-resolution satellite imagery, here we consider including landscape features from satellite imagery into infectious disease prediction models. To do so, we implement a Convolutional Neural Network (CNN) model trained on Imagenet data and labelled landscape features in satellite data from London. We then incorporate landscape features from satellite image data from Pakistan, labelled using the CNN, in a well-known Susceptible-Infectious-Recovered epidemic model, alongside dengue case data from 2012-2016 in Pakistan. We study improvement of the prediction model for each of the individual landscape features, and assess the feasibility of using image labels from a different place. We find that incorporating satellite-derived landscape features can improve prediction of outbreaks, which is important for proactive and strategic surveillance and control programmes.
\end{abstract}

%
%

\maketitle

\input{main_input}

\bibliographystyle{ACM-Reference-Format}

\bibliography{eg_bib}

\end{document}

%% file: main_input.tex
\section{Introduction}

Vector-borne diseases cause more than 700,000 deaths every year globally. Mosquitoes are the best known disease vector, others include ticks, flies, and fleas \cite{VectorBorneWHO}. Dengue virus is the most ubiquitous mosquito-transmitted disease, and is transmitted primarily by Aedes aegypti mosquitoes, a vector which also transmits Zika, chikungunya and yellow fever \cite{simmons2012dengue}. The virus disproportionately affects urban areas in developing countries, which often have limited resources for containment and intervention activities \cite{halstead1984selective}.

Spatial patterns of land use and land cover play an important role in infectious disease dynamics. For example, landscape alterations, such as residential developments or agricultural land use, may result in higher contact rates between humans and disease vectors \cite{smith2005ecological}. Other types of land cover may relate to disease prevalence as they can create favorable conditions for vectors and/or hosts. On the other hand, human-induced landscape alterations may induce habitat loss, species extinction, altered nutrient dynamics, invasive species colonization, and other ecosystem processes leading to a change in disease incidence rates \cite{jackson2006towards}. 

Remote sensing data is frequently utilized for land use/land cover classification. Organizations such as the National Oceanic and Atmospheric Administration and the U.S. Geological Survey also publish landcover data, which generally use a mix of spectral change for classification, automated and expert classification of satellite data \cite{wickham2018spatial}. The resolution of such data varies, but typically can range from $\sim$10's to 100's of meters \cite{landsatweb}. Moreover, landscape features in relation to disease incidence, have generally been related to disease through regression models, and do not account for underlying non-linear relationships between environmental features and disease incidence, nor the partial nature of the observations.

More recently, studies using Google Maps images take advantage of higher resolution data from satellites such as DigitalGlobe (0.15-1.24m resolution). Such publicly available remote sensing data has recently been harnessed for many social applications. From monitoring wheat fungus in crops \cite{pryzant2017monitoring}, to predicting poverty \cite{jean2016combining}, crop yields \cite{wang2018deep} and correlating with obesity \cite{maharana2018use}. The higher resolution potentially provides the opportunity to distinguish various landscape features, such as boundaries of homes and roads, and presence of individual trees, which can have a potential impact on the transmission of diseases.

In this paper, we present a deep learning model for acquisition of deep landscape features from high-dimensional satellite imagery, and also demonstrate integration of these deep features in an infectious disease prediction model and assess their performance for improving infectious disease prediction. To overcome the lack of training data on Pakistan images, we use a transfer learning approach. We also assess the added value of each landscape feature. Finally, given the expected limitation(s) of using training data from another location, we study how the transfer performs in urban versus rural regions of the Pakistani cities. Our results show that landscape features derived from automatically learned features improve prediction of dengue case data (even with transferring information from landscape features from another location). This work serves as a valuable proof of concept for integration of new data sources into infectious disease modeling, and directions for future work. Given that the satellite data is freely-available, this work is easily scalable, which is especially relevant for places affected by vector borne diseases. 



\section{Related Work}
Landscape features, collected at various spatial and temporal resolutions, have been used in the past to identify disease risk. Studies which either use pre-labelled landcover maps, primitive image segmentation or classification approaches to label satellite imagery include Vanwambeke et. al, who analyzed how changes in landscape features such as forests, orchards and dam construction changes amount of malaria and dengue spreading mosquitoes in Thailand. The study uses coarse-grained Landstat data collected at a resolution of 30m \cite{vanwambeke2007impact}. Another work by Vanwambeke et. al also using low resolution satellite imagery data, studied the relationship between abundance of Aedes and Anopheles larvae, and 5 land cover features: forest, irrigated fields, orchards, peri-urban housing and villages \cite{vanwambeke2007landscape}. Nakhapakorn et. al classified low resolution satellite imagery data from Landsat, into 4 classes: build-up, water, agriculture and forest areas using Maximum Likelihood Classification, and identified the percentage of dengue incidence occurring in each class in a province in Thailand \cite{nakhapakorn2005information}. Use of low resolution satellite imagery or pre-labelled maps hinders both the use of high resolution landscape features, and scalability of the approach in other parts of world. In addition, these studies have modelled landscape features as part of linear models which often do not account for the underlying dynamics of disease, as opposed to disease transmission models.

Some studies have used vegetation metrics, such as enhanced vegetation index (EVI) and normalized difference vegetation index (NDVI), extracted from satellite images, to model dengue incidence \cite{23,2}. While these metrics have shown to correlate with dengue risk, they do not enable measurement of non-vegetation landscape features such as building and water bodies. Similarly, nightlight imagery has been used as a proxy of changes in populations to model measles in Niger \cite{bharti2011explaining}. This study does not use any framework to first extract features from the nightlight images, and instead directly uses averaged pixel intensity values.

More recently, studies have started using deep learning frameworks to classify satellite images for disease and related predictions. The high dimensional output of a VGG neural network, as individual features, was used to predict the prevalence of obesity in a neighborhood \cite{maharana2018use}. Satellite imagery data has also been used to predict poverty levels in five countries in Africa. The study uses a VGG architecture to infer ‘representation’ of nightlight images, from corresponding daylight images, and use them to predict poverty \cite{jean2016combining}. While these studies perform well, the lack of knowledge about which landscape features are represented through the high dimensional output of a neural network, makes it difficult to assess the relationship between individual landscape features and the problem under consideration.

\section{Datasets}
Here we use publicly available high resolution satellite imagery data from Google Maps API. Using boundaries for the cities of Lahore and Rawalpindi in Pakistan, we download satellite imagery data at a zoom level of 17 with each image scaled to a factor of 2 ($n=$8,632 images 6,476 for Lahore, 2,156 for Rawalpindi). Each downloaded image is of native resolution 1280 by 1280 pixels.

Given the absence of semantic segmentation maps for the city of Pakistan, we use labeled data from the Kaggle DSTL (Defence Science and Technology Laboratory) competition (https://www.kaggle.com/c/dstl-satellite-imagery-feature-detection) for training purposes. The datasets consists of 25 images of resolution 3600 by 3600. The dataset also contains segmentation maps (pixel-wise labels) for 10 labeled classes corresponding to each image. We 
selected landscape classes for which there is some prior knowledge regarding mechanistic relevance to vector-borne disease spread. Roads, buildings and crops were included given that the movement of individuals and urbanicity in a location impacts dengue transmission \cite{32}. Additionly, trees and water bodies like standing water and waterways provide good breeding sites for dengue transmitting mosquitoes \cite{47} and hence were also included. 


Confirmed dengue incidence data from both cities was received from the Punjab Information Technology Board. Details of confirmed dengue incidence from all public hospitals in Pakistan's Punjab province are aggregated in a centralized server. 
The dataset consisted of 10,888 cases reported in the cities of Rawalpindi ($n$=7,890 between January 1, 2014 and December 31, 2017) and Lahore ($n$=2,998 between January 1, 2012 and December 31, 2017). City-wide daily mean temperature and precipitation were obtained from the Pakistan Meteorological Department \cite{PMD}. High resolution population data was retrieved from WorldPop \cite{worldpop}, and previously published work \cite{28}.

\section{Methods}

\begin{figure}[h]
	\begin{center}
		\includegraphics[width=0.99\linewidth]{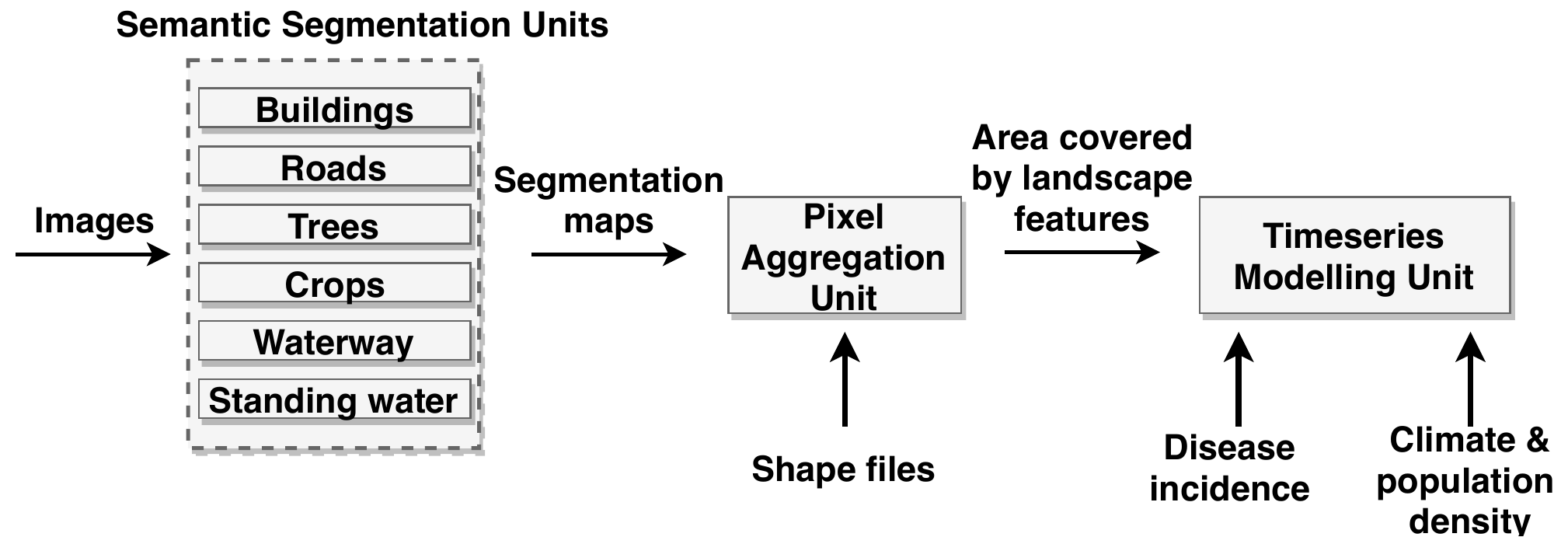}
	\end{center}
	\caption{Pipeline of the multi-step methodology used in the study.}
	\label{fig:pipeline}
\end{figure}

\subsection{Overview}
We use a multi-step approach to understand if the addition of landscape features improves the prediction of dengue transmission in the cities of Lahore and Rawalpindi in Pakistan. To achieve this, we first use a deep learning approach to create segmentation maps for the six landscape classes, from satellite images from both cities. Given the localized nature of dengue transmission, it is helpful to model the transmission of dengue at a sub-city spatial resolution in each city. Hence the segmentation maps of each landscape class, extracted from the deep learning approach, are aggregated at a sub-city spatial unit level in each city to identify the percentage of a spatial unit covered by a particular landscape class. The landscape features are then included in a common time series disease modelling framework that accounts for population mixing, disease specific properties and exogenous parameters and the processes underlying relations between those factors and disease incidence \cite{17,23}. In addition to the extracted landscape features, weather parameters and population density, which are commonly used, are included to model the transmission of dengue and predict disease incidence over time. Figure \ref{fig:pipeline} shows the pipeline of the multi-step approach used here.

\subsection{Semantic Segmentation}
\subsubsection{Architecture}

\begin{figure*}[h]
	\begin{center}
		\includegraphics[width=0.95\linewidth]{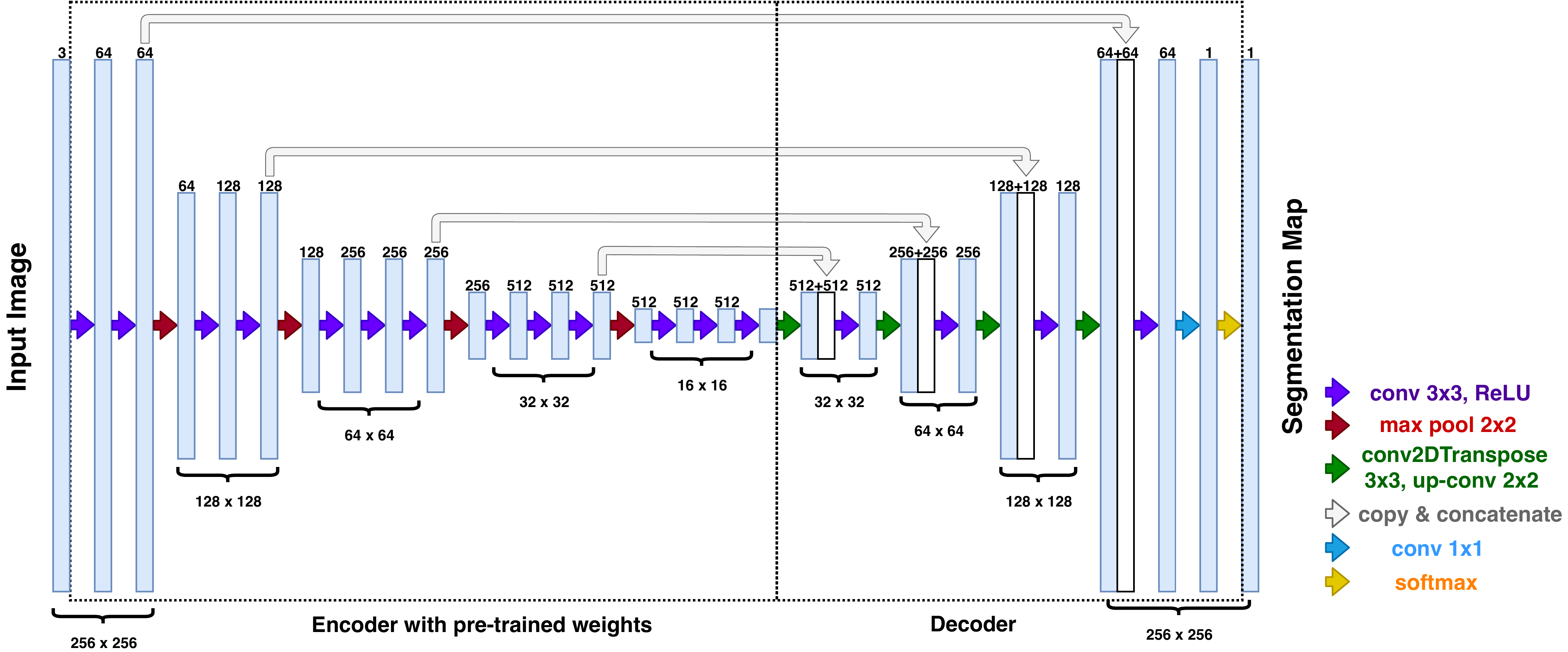}
	\end{center}
	\caption{Architecture of U-net used for semantic segmentation. Dotted lines enclose the encoder and decoder. The encoder consists of 13 3x3 convolutions each followed by a rectified linear unit (ReLU), and 4 2x2 max pooling operations with strides of 2 to reduce the map size by 2. Weights of all layers in the encoder are pre-trained on ImageNet dataset. The decoder consists of 4 3x3 convolutions each followed by a rectified linear activation unit (ReLU) and 1 1x1 convolution. The decoder also consists of 4 3x3 transposed convolutions with strides of 2 to upsample maps by 2. The upsampled maps are concatenated with skip connections from the encoder. The architecture has a final softmax activation which predicts the probability of each pixel in the image belonging to the landscape class.}
	\label{fig:architecture}
\end{figure*}

To generate segmentation maps (pixel-wise labels) of individuals landscape features from satellite imagery, we use a U-Net architecture, a class of CNN, given they tend to perform well for semantic segmentation tasks with limited training data \cite{ronneberger2015u}. The architecture consists of a series of successive convolutional and maximum pooling layers, followed by a series of convolutional and up-convolutional layers (Figure \ref{fig:architecture}). 
The encoder extracts features, while the decoder reconstructs dense representation of pixel-wise classification using the extracted features from the encoder. Skip connections, present in a U-net architecture from the encoder to decoder, allow incorporation of both global and local features present inside an image in the final output to improve the pixel-wise classification \cite{ronneberger2015u}.

Given that both the encoder of U-Net and VGG neural networks consist of a series of convolutional and max pooling layers, and given the lack of a large training dataset available to train the neural network from scratch, we use a VGG16 neural network, trained on ImageNet data, as the encoder of the architecture. A similar methodology, but with VGG11 architecture as an encoder has been described elsewhere \cite{iglovikov2018ternausnet}.  ImageNet data consists of millions of labelled images from over a thousand categories \cite{russakovsky2015imagenet}, and networks trained on this dataset are known to be good extractors of visual features such as edges and corners \cite{jean2016combining}. The fully connected layers at the end of the VGG16 architecture are discarded, given they represent the decoder part of the VGG architecture. The remaining convolutional layers are then used as a pre-trained encoder. The decoder, to complement this pre-trained encoder, is then constructed with randomly initialized weights. 

\subsubsection{Training, optimization and prediction}
We train separate networks for each landscape feature, given that landscape classes in the training dataset are not mutually exclusive \cite{iglovikov2017satellite}. We use an input dimension of 256 by 256 by 3 pixels to the network, where the third dimension represent the red green blue (RGB) band of an image. The input size is chosen to ensure that the network does not cause memory issues when training and is comparable to that used in other work harnessing similar features from satellite imagery \cite{xie2016transfer}. In addition the size is chosen as a multiple of 1024 to ensure that the images, to be predicted, can be subset into non-overlapping sub images without padding white spaces on boundary images \cite{ronneberger2015u}. The output of the network is 256 by 256 by 1 representing the probability of each pixel belonging to a particular landscape class for which the model is trained. We use an Adam optimizer with the default learning rate of 0.001 and optimize the loss function $C$, which is defined as: $C = B - log(J)$, where $B$ is the binary entropy between predicted and actual segmentation maps and $J$ is the Jaccard similarity index between predicted and actual segmentation maps. 

To learn the weights of the decoder we use data from the Kaggle DSTL competition. Pixels in each image in the dataset are normalized using the combined mean and the standard deviation of the pixels in the training dataset. The images and corresponding pixel-wise labels in the dataset are then partitioned into non-overlapping blocks of sub-images of size equal to the input size of the network. The network is then trained using the sub-images.  



To generate pixel-wise labels for images in Lahore and Rawalpindi, we first re-size all images from Google API in our set to 1024 by 1024 pixels. Pixels in the images are then normalized using the combined mean and the standard deviation of pixels in the downloaded images from Google API, as opposed to images from the DSTL challenge. This ensures that any differences, due to variation in exposure or lightning, when capturing the DSTL and Google Maps images, are minimized.  Each image is then partitioned into non-overlapping sub-images of the size of input of the network, and predictions made using the pretrained models for each landscape feature. Keras library in Python, with a Tensorflow backend is used for semantic segmentation.

\subsection{Timeseries Modelling}
We model the dengue transmission using a timeseries susceptible infected recovered (TSIR) model of viral incidence which has been widely used in epidemiology to reconstruct dynamics of diseases \cite{17,23}. Each city is divided into sub-city spatial units (towns) to model localized transmission ($n$=10 spatial units in Lahore and $n$=14 spatial units in Rawalpindi). Spatial units for the city of Lahore are the sub-city resolution administrative boundaries, while for the city of Rawalpindi we use boundaries as defined by Rehman \textit{et.al} \cite{rehman2018quantifying} based on clusters of dengue incidence in the city.

Given that only the most recent satellite imagery data was available, we model landscape features as time-invariant in the dengue transmission model. We calculate the proportion of area, in each spatial unit, covered by each landscape feature. To achieve this, we first select the images belonging to each spatial unit using the boundaries of the spatial units. The predicted pixel-wise labels of the selected images are then used to calculate the ratio of total number of pixels representing the presence of a landscape feature, and the total number pixels in the selected images.

\begin{figure*}[t!]
	\begin{center}
		\includegraphics[width=0.99\linewidth]{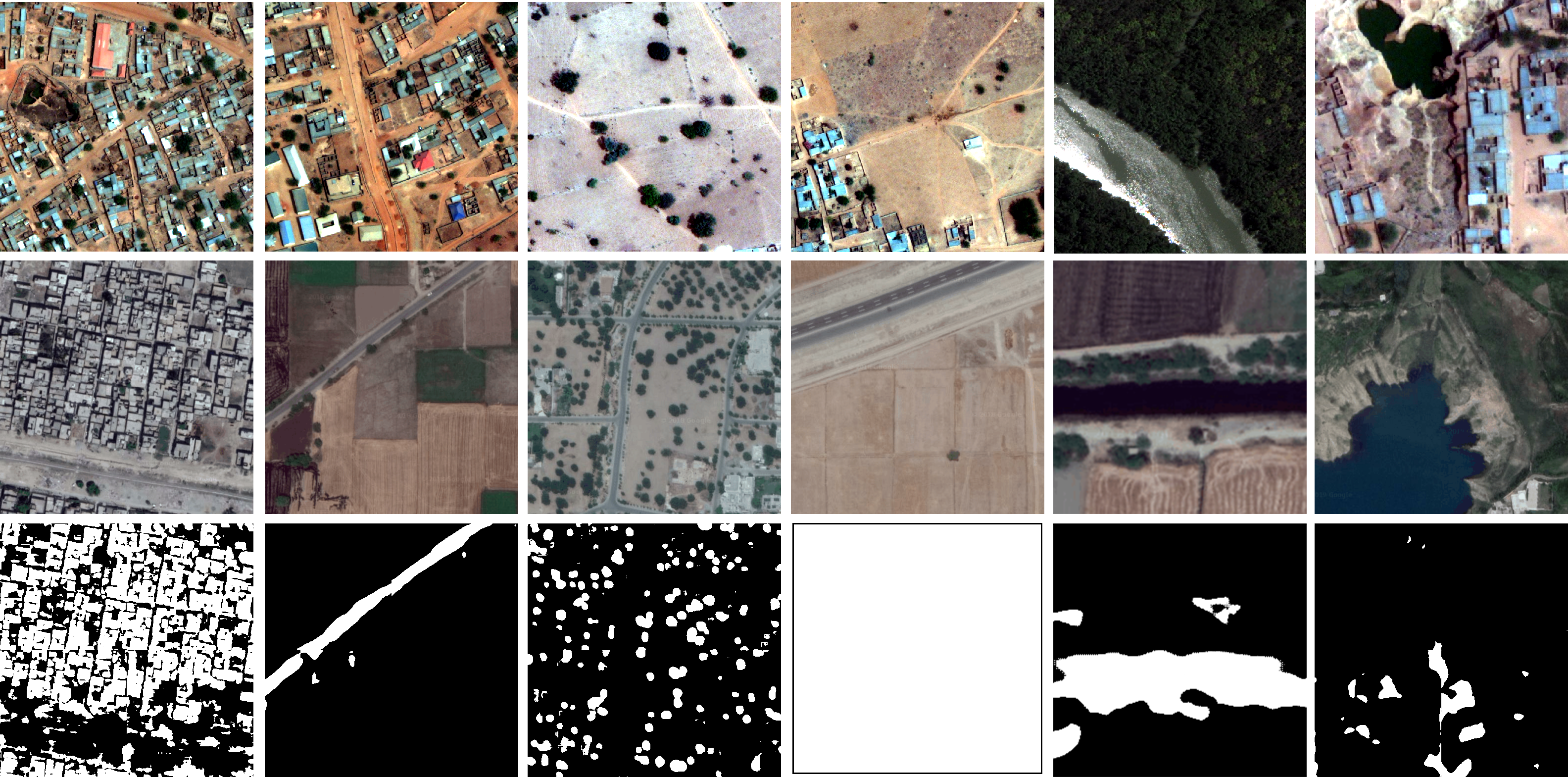}
	\end{center}
	\caption{Sample images of landscape features present in DSTL data (row 1) and Google Maps API data (row 2), and predicted segmentation maps corresponding to the Google Maps API data (row 3). Landscape features from left to right: building, roads, trees, crops, waterway and standing water. Higher intensity corresponds to a higher probability of a pixel belonging to a landscape class.}
	\label{fig:pred_maps}
\end{figure*}

The entire modeling approach is described below succinctly, however we emphasize that such approaches, parameters chosen, etc. are the standard in infectious disease modeling. Further details on this modelling approach can be found in Rehman \textit{et.al} 's work \cite{rehman2018quantifying} from where the specific dengue TSIR modeling methodology has been adapted. The general TSIR model, for each spatial unit, $i$, is defined via:

\begin{equation}
	I_i(t+1) = \beta_i(t) \frac{S_i(t)}{N_i(t)}I_i^{\alpha_i}(t) \epsilon
	\label{ts:eq1}
\end{equation}

\noindent where an interval of 2-weeks is used for $t$. $I_i(t)$, $S_i(t)$, and $N_i(t)$ are the infected, susceptible and total population during time step $t$ in spatial unit $i$, $\alpha_i$ is the mixing coefficient in $i$, and $\beta_i(t)$ is the transmission rate during time step $t$. The error term $\epsilon$ is assumed to be an independent and identically log-normally distributed random variable. Weather parameters (bi-weekly mean temperature and number of rainfall days) and population density are commonly incorporated in such models given they are known to affect dengue transmission \cite{xu2017climate,world2009dengue}. Time invariant landscape features, and time-variant weather parameters and population density are modelled as part of the transmission rate. Equation \ref{ts:eq1} is rearranged as:
\begin{equation}
	log(\beta_i(t)) + \alpha_i*log(I_i(t)) = \\
	log(I_i(t+1)) + log(N_i(t)) - log(S_i(t)) - \epsilon
	\label{ts:beta}
\end{equation}
and the transmission rate is substituted with:
\begin{equation}
	log(\beta_i(t)) = \\
	\sum_a{\theta_a L_{i,a}} + \sum_j{\theta_j E_{j}(t-l_j) + \theta_p D_i(t)}
	\label{ts:eq3}
\end{equation}
where $l_j$ are the delays in time steps which are added to weather parameters $j$ to account for vector life cycle. $L_{i,a}$ is the percentage of area of spatial unit $i$ covered by landscape feature $a$. $E_{j}(t-l_j)$ is the value of weather parameter $j$ during time step $(t-l_j)$. $D_i(t)$ is the population density in spatial unit $i$, which is known to also drive transmission, $\beta_i$, and is computed by dividing the population of the spatial unit with the area of the spatial unit. {$\theta_a$, $\theta_j$ and $\theta_p$} are parameters which relate landscape features, weather and population density to $\beta$. 

We use a linear model to fit the relationship in equation \ref{ts:beta} and then study how the fit varies for model with and without the addition of landscape features. In addition, we also study the fit of the model by incorporating each type of resolved landscape feature individually in the model.

\begin{table*}[h]
	\begin{center}
		
		\begin{tabular}{l|lll|lll}
			\hline
			Model                         & \multicolumn{3}{c|}{Lahore}          & \multicolumn{3}{c}{Rawalpindi}      \\
			\hline
			& all towns & more urban & less urban & all towns & more urban & less urban \\
			\cline{2-7}
			Environment only              & 0.728 & 0.736 & 0.715 & 0.774 & 0.801 & 0.660  \\
			All landscape \& environment   & \textbf{0.747} & \textbf{0.757} & \textbf{0.734} & \textbf{0.808} & \textbf{0.832} & \textbf{0.717} \\
			\hline
			Building \& environment        & 0.732 & 0.746 & 0.725 & 0.784 & 0.824 & 0.693 \\
			Road \& environment            & 0.731 & 0.746 & 0.719 & 0.781 & 0.804 & 0.678 \\
			Trees \& environment           & 0.730  & 0.742 & 0.725 & 0.785 & 0.813 & 0.678 \\
			Crops \& environment           & 0.728 & 0.736 & 0.715 & 0.774 & 0.801 & 0.660  \\
			Waterway \& environment        & 0.737 & 0.745 & 0.725 & 0.796 & 0.817 & 0.702 \\
			Standing water \& environment & 0.730  & 0.736 & 0.723 & 0.778 & 0.808 & 0.663 \\
			\hline
		\end{tabular}
	\end{center}
	\caption{Adjusted R-squared values of model fit using only i) weather and population density (environment parameters), ii) one landscape feature and environment parameters, and iii) all landscape features and environment parameters. The model fit values are shown across i) all towns, ii) more urban towns, and iii) less urban towns. Best model results in bold.}
	\label{tab:results}
\end{table*}

\section{Results}
First, we use a transfer learning approach to generate segmentation maps of six landscape features from the satellite imagery data of the cities. The trained architectures for landscape classes show good performance when predicting the segmentation maps on the 20\% held-out labelled data from the DSTL challenge. Specifically, the architecture provided a higher Jaccard similarity for buildings (0.628), roads (0.660), trees (0.557), crops (0.813), and waterway (0.618). Performance on standing water was lower (0.357). The estimated Jaccard index values on the held out dataset are slightly lower than those reported in a previous work which uses DSTL data \cite{iglovikov2017satellite}, but in our work we only use the RGB bands to train the architectures, and not the additional 8 multi-spectral and 1 short-wave infrared bands as we will be predicting on satellite image which only use RGB bands. 

Though the encoder of the architectures were pre-trained on ImageNet data, and decoder on DSTL challenge data from London, our goal was to eventually make predictions on data from Pakistan. Visually analyzing predicted segmentation maps for satellite imagery data in cities of Pakistan (Figure \ref{fig:pred_maps}) 
shows that despite using no labelled data from Pakistan, the architectures were able to identify the segmentation maps of buildings, roads and trees with a reasonable quality. Results for waterways provided medium quality. Identification of crops and standing water landscape features on Pakistan data were not meaningful.


Epidemic model results which included proportion landscape features, derived from the satellite imagery, by area result in time series of predicted dengue cases. For evaluation we use the adjusted $R^2$ metric, which allows the comparison of the fit of various models while accounting for the number of parameters being used in the model. The base model trained only on weather parameters and population density provide a good fit with the incidence of dengue, with an adjusted $R^2$ value of 0.728 in Lahore and 0.774 in Rawalpindi. Addition of the landscape features in the model improved the fit (0.747 in Lahore and 0.808 in Rawalpindi). 

To examine how the London training data generalized to different types of Pakistan landscapes, we assessed the goodness of fit in more versus less urban spatial units in both cities, as this is a natural potential difference between the two places. For each city, we identified spatial units which have more areas covered by buildings compared to the average area covered by buildings in the spatial units. We found that the fit of model is better in more urban areas as compared to less urban areas, across both cities. (0.757 as compared to 0.734 in Lahore and 0.832 as compared to 0.717 in Rawalpindi). Incorporation of landscape features individually showed varied improvement in the fit of the model. Overall we find incorporation of buildings and waterways to provide the most improvement in results, followed by roads and trees. Incorporation of crops and standing water provided little or no improvement in the fit. Table \ref{tab:results} summarizes the results across all models in both cities.

\section{Conclusion and Future Work}
Here we use a multistep approach to study the improvement in disease prediction by incorporation of landscape features extracted from satellite imagery. 
To extract landscape features we use a deep learning architecture trained on ImageNet and DSTL challenge data from London. Landscape features are then modelled as part of a state-of-the-art TSIR disease modeling framework, to study the improvement in fit of the model representing disease transmission. 

Despite differences in the types of roofs and color of roads in London and cities in Pakistan, architectures trained for buildings and roads were able to predict the segmentation maps for satellite data in Pakistan with a reasonable accuracy (Figure \ref{fig:pred_maps}). Predictions for trees were the most accurate amongst all landscape features, given the consistency of color and shape of trees across the datasets. Given the variation in waterways in both datasets, despite relatively high Jaccard index for this feature, the trained architecture was only able to identify approximate location of waterways and not the exact outline on the Pakistan data. 
Results in Table \ref{tab:results} reflect not only the predictive power of the landscape features, but also how correct the generated segmentation maps were for the landscape features. Little to no improvement in the adjusted $R^2$ value with the addition of crops and standing water in the model does not conclude that these features are not predictive of dengue transmission.  Lack of predictive performance in the model when including standing water feature can be attributed to the already low performance of the architecture on the held out training data. In contrast, the deep learning architecture trained for crops provided the highest predictive performance on held out data, yet given the lack of variation in shapes and color of crops in the training data, the architecture was not able to learn generalizable features for transfer to the Pakistan data; when applied to satellite imagery data from Pakistan, the model tended to predict every image as entirely consisting of crops (example subplot in Figure \ref{fig:pred_maps} consisting of all white pixels). Amongst all landscape features, buildings provided both a large improvement in the fit of the timeseries model and segmentation maps which were reasonably accurate.

Results consistently showed that addition of landscape features improved the fit of the model across both cities, and in both rural and urban areas. Base models for the cities provided a better fit in urban areas as compared to rural areas as dengue is more prevalent in urban areas. In addition models for the city of Rawalpindi provided a better fit compared to those for Lahore, given that the disease activity was higher in Rawalpindi during the study time period.

While results demonstrate the utility of adding landscape features in timeseries modelling of dengue, steps can be taken to make the results more robust. First, using labelled satellite imagery data from Pakistan can help identify the segmentation maps for Pakistan-specific features and improve predictions for the remaining features. Second, given the high spatial resolution of satellite imagery data, modelling the transmission of dengue at an even higher resolution could potentially help identify more robust relationships between landscape features and disease transmission. Also, we only used satellite images from one time period in this study, based on availability. This may be fair as several considered landscape features may not have changed over the time period of the study. However for features where there is expected differences, it may be useful to resolve them with more temporal resolution. Finally, an unsupervised approach, using the high dimensional output of the neural network, as done in previous work \cite{maharana2018use, jean2016combining}, may identify unknown features which can potentially impact disease transmission.